\documentclass{article}
\usepackage[latin1]{inputenc}
\usepackage[dvips]{graphicx,epsfig,color}
\usepackage{wrapfig,rotating}
\usepackage{amssymb,amsmath,array}
\usepackage{cite}
\def\Journal#1#2#3#4{{#1} {\bf #2}, #3 (#4)}

\def\NPB{{\em Nucl. Phys.} B}
\def\PLB{{\em Phys. Lett.}  B}
\def\PRL{\em Phys. Rev. Lett.}
\def\PRD{{\em Phys. Rev.} D}

\def\beq{\begin{equation}}
\def\eeq{\end{equation}}
\def\bea{\begin{eqnarray}}
\def\eea{\end{eqnarray}}

\begin{document}

\begin{flushright}
Freiburg-PHENO-2010-005.
\end{flushright}
\vskip 1.0cm
\begin{center}
{\Large \bf Gravitational anomaly and fundamental forces}
\vskip 0.4cm
J.~J.~van~der~Bij\\
\vskip 0.3cm
Insitut f\"ur Physik, Albert-Ludwigs Universit\"at Freiburg \\
H. Herderstr. 3, 79104 Freiburg i.B., Deutschland\\
\end{center}
\vskip 0.7cm

\begin{abstract}
I present an argument, based on the topology of the universe, why there
are three generations of fermions. The argument implies a preferred unified gauge
group of $SU(5)$, but with $SO(10)$ representations of the fermions.
The breaking pattern $SU(5) \rightarrow SU(3)\times SU(2)\times U(1)$
is preferred over the pattern $SU(5) \rightarrow SU(4)\times U(1)$.
On the basis of the argument one expects an asymmetry in the early universe
microwave data, which might have been detected already.
\end{abstract}
\vskip 0.7cm
\section{Introduction}
As the title of this contribution indicates I will be concerned with the
fundamental forces of nature. As fundamental forces I consider the gauge forces
of nature. Included are also the fermions that form representations of the gauge groups.
I will in first instance not be concerned with the Higgs mechanism and the breaking
of the gauge symmetries. The world described therefore consists in this picture of
massless fields only. One can legitimately ask why one would study such a system and why it
should be a subject for a school on Quantum Gravity and Quantum Geometry.
There is a twofold reason why such a study could be interesting, one coming from
particle physics phenomenology, the other being related to quantum gravity.

When one considers the status of particle physics phenomenology it is generally stated that the
standard model gives an excellent description of the data, but that the model is 
unsatisfactory with respect to the Higgs sector. Emphasis is put on the so-called 
naturalness or hierarchy problem, which consists of the statement that the Higgs mass
receives quadratically divergent contributions, so that the Higgs mass should be of
the order of
the cut-off scale for which there is no evidence in the data. Great efforts are subsequently made
to solve this hierarchy problem. One introduces various extensions like supersymmetry,
technicolor, higher dimensions and so forth. All these extensions have problems with the data
and one has to finely tune many parameters or invoke dynamical accidents in order not 
to get into conflict with experiment. I think it is fair to say that this approach has lead to an impasse.
One should therefore question the basic assumption that the most important subject to study
is the symmetry-beaking of the theory.  At the moment we have no clue why the symmetry
of nature is described by the gauge groups that are seen in experiment, in particular
also why there are exactly three generations. Maybe one should first understand this
question why we have the symmetry that we see in experiment, before one should try to
understand its breaking. The structure of the symmetry might involve deeper principles,
that ultimately will be reflected in the symmetry breaking too. At our present state of knowledge
worrying about the hierarchy problem is then simply premature.

A second reason to study this system can be found in the status of quantum gravity.
At present there are two leading proposals, or rather classes of proposals, that
might lead to a fundamental theory of gravity, string theory and canonical gravity.
Most of the effort in the last decades has been spent on string theory.
The reason is apparently, that string theory promises to be not only a theory of
gravity alone, but would also include matter fields. Thereby one would arrive at a 
fully unified theory of all forces and in the best of all worlds a rather unique one.
With the arrival of the landscape the uniqueness has disappeared and thereby string theory
cannot explain why the gauge forces are the way they are. Unification within the context
of canonical quantum gravity has received little interest, because  at first sight there 
is no connection. However this is where the term anomaly in the title comes in.
There may be conditions on representations of gauge fields coming from anomalies,
that could play a role. If these restrictios are strong enough one could argue that 
canonical quantum gravity is the way to a unification of forces.

So basically we therefore want to see whether there is a uniqueness to the
fundamental forces and particles that exist in nature.
Is there a reason for the gauge group and representations of nature?
 In particular we would like
to understand why there are three generations of fermions. These are rather 
old questions. The generation question was asked by I. Rabi after the discovery
of the muon: "Who ordered that?" The question about the uniqueness of nature's laws
was paraphrased by A. Einstein: "Did God have a choice when he created the world?"
The argument\cite{me} that I present in the following will give
 tentative answers to these questions.

\section{Ancient history}
As is clear from the above we are trying to derive the fundamental forces and particles
of nature from first principles. This is a rather ambitious project and one might
be accused of hubris and look silly in the end. In order to try to prevent this from happening
we will look at similar attempts in the past and see if we might learn from these.
We know of three attempts.
The first takes us to the ancient Greeks. The earliest Greek philosophers had a problem
regarding nature, that is hard to understand for modern people. The problem had to do
with the question of motion and the unity of nature. It was related to
the concepts of emptiness and fullness. If the world is empty there is nothing to move.
If the world is full there is no place to move. Still motion exists. The solution was
constructed by Leukippos and Demokritos. They split reality in two separate pieces,
on the one hand space which is empty, on the other hand the atoms that are full.
Movement was then explained as atoms moving through empty space. Such atoms we would nowadays
probably call molecules. 
A second line of research was developed by Plato, Empedokles and Aristoteles, probably
under the influence of Pythagorean thinking, which introduced mathematics in the 
description of nature. It was concerned with the elements, which we would consider the
basic building blocks of the atoms. One considered four elements and associated 
regular bodies to them: fire (tetrahedron), air (octahedron), water (icosahedron)
and earth (cube). There was some obvious phenomenology associated to this, for instance
fire has to be a tetrahedron, because it has sharp points, so one burns oneself.
This theory has a nice mathematical structure based on symmetry and actually made a prediction
that there should be a fifth element corresponding to the dodekahedron, known as
quintessence. It was variously identified as the ether, the soul and so forth.
Present theories of quintessence are roughly at the same level.
With hindsight we might consider this all to be nonsensical, but that  is wrong.
This was one of the most succesful theories for physics and chemistry, being the
leading theory for about a thousand years. It did not disappear, because it is
was manifest nonsense, which it is not, but because of alchemical experiments
and experiences  in medicine, which made the theory untenable.

\section{Modern history}
We now skip  a few thousand years and ask: where do we stand?
Of course we understand better what physical laws are, basically differential
equations, but at the philosophical level regarding the basic structure
of reality things are still rather similar.
We now have spacetime instead of space,
which is not such a great leap philosophically speaking.
More important however is that spacetime is not anymore static, it is a dynamical
quantity, that is related to matter through the Einstein equations.
Thereby the split of reality into two parts, space and atoms, has been partly closed.

Regarding the structure of matter things are very similar. Symmetry still determines
the elements. Instead of Platonic solids, determined by symmetry, we now have gauge groups
and representations that are determined by symmetry. In a sense we have made a step backwards,
since there appear to be an infinite number of possibilities for the groups and
representations. The only constraints are the ones coming from the known
(chiral) anomalies.

Given the fact that the Einstein equations connect matter and spacetime
it is a natural attempt in unification to connect the two via a deeper principle.
One takes the largest, most symmetric (beautiful) theory one can find and postulates
that nature should be described by this theory. In order to do this one normally 
has to assume a unification of the gauge forces, the simplest being the unification
to the group $SU(5)$.
  
 An early attempt to derive 
$SU(5)$ was made in the context of $N=8$ supergravity \cite{ellis}.
In order to derive a model that approximates nature, a  number of dynamical
assumptions regarding composite states had to be made.
With the realization that $N=8$ supergravity is not a
finite theory and therefore not suitable as a fundamental theory
for all forces, attempts along this direction have largely stopped.

A somewhat later attempt was to use string theory in the form of the 
heterotic string, which implies a gauge group
 $E(8)\times E(8)$
\cite{rohm}.
 An argument similar to the one
we present later was used. The group is selected by the absence
of an anomaly, in this case the conformal anomaly 
on the world sheet. Subsequently reducing the group to
something closer to the standard model appeared possible, but rather 
complicated. The number of fermion generations is determined by
topological considerations. With the realization, that string theory allows for many vacua
with different gauge groups, the idea of a unique group has been abandoned
in this approach. These two attempts are similar in that they are very ambitious.
The assumption is that one determines the unique form of fundamental dynamics
from a given mathematical structure, which should subsequently contain
the observed forces of nature.

What do we learn from these earlier attempts? 
The first thing we should learn is some modesty.
Assuming that one  knows the fundamental laws of phyics
and only has to construct the standard model out of these,
is not a very promising approach. Mankind is not smart enough
for that. Therefore we have to use experimental information.
Furthermore we found that anomalies are important. As a related point
also topology appears to play a role. Fortunately there are new results since 1985,
 in phenomenology, cosmology and mathematical
physics, that under a suitable interpretation might help us to move
forward. Of course this involves a certain guesswork, regarding which features
are  important.

\section{Phenomenology}
If one wants to construct a fundamental theory of nature,
the idea of unification appears necessary. 
The standard model based on the gauge group $SU(3)\times SU(2) \times U(1)$
with its complicated set of fermions, constrained by the anomaly is
too peculiar to be fundamental. The situation is
simply asking for a unification into a larger group. It is well known
that such a unification is possible and quite natural. 
When one considers just the gauge bosons of theory the simplest form of unification
is within the gauge group $SU(5)$. Important in this respect is that the rank of the
group is four. Even though the symmetry has to be broken, one would 
in first instance expect the rank of the broken group to be the same
as of the unbroken group. The question is therefore if there is any evidence in the data
for a higher rank of the gauge group.

This question has been studied at the LEP experiments.
Various analyses are possible. 
The upshot is that there is no convincing evidence for the existence of
extra $Z's$ or $W's$. Therefore the known vector bosons point towards a unification
within a group of rank four, namely $SU(5)$ \cite{georgi}.

In the fermion sector the situation is somewhat different.
At the latest with the discovery of neutrino masses, it has become clear
that the natural unification for the fermions is within the group 
$SO(10)$ \cite{fritzsch},
since each generation forms an irreducible spinor representation of $SO(10)$.
Moreover the spinor representation of $SO(10)$ is somewhat special, since it is the smallest
triangle anomaly free complex representation in all of group theory\cite{bert}.

So naively speaking the vector bosons and the fermions point toward a different
form of unification. Of course the situation can be described through the breaking
of the symmetry with a number of Higgs fields, but one would hope for
a more fundamental explanation for this feature. 
As $SU(5)$ is a subgroup of $SO(10)$ there ought to be an {\bf obstruction}
to gauging the full group $SO(10)$.

Another fact of phenomenology is
the existence of precisely three generations of fermions. 
This is established from the invisible decay width of the $Z$ boson and 
by the precision measurements at LEP.
It is natural to wonder
whether the group question $SU(5)$ versus $SO(10)$  is related to  the 
question of the number of generations. So  we have to explain:
why $SU(5)$ for vectorbosons, why $SO(10)$ for fermions, why $3$ generations?
We are therefore looking for an argument to constrain the representation content
and the gauge group of the theory. The only type of argument known that can give
such constraints is based on some form of an anomaly. As anomalies are intimately 
related to topology, one is led to the question: what is the topology
of space?\\

\section{Topology in cosmology}
The question of the topology of the universe has many aspects. 
It has been studied in great detail in Kaluza-Klein theories, where a
typical assumption is of the form $M_4 \times S_1^n$, a torus shape.
The higher dimensions form circles, of which one hopes that the radii  shrink to zero
as the universe develops. It has appeared rather difficult to realize this picture
in an attractive way in practice.
We therefore take the opposite point of view and assume that the universe is
lower dimensional, specifically the spatial part is supposed to be two-dimensional
instead of three-dimensional. At first sight this statement appears to be obvious nonsense 
and would earn the author
a prize from the flat earth society, if not further qualified.
 Of course what is meant here is that the universe
is two-dimensional at the origin of the universe and the third dimension becomes large
only at late times. The idea is that the universe has the topology $M_3 \times S_1$, where the
radius of the circle shrinks to zero when one goes back in time.
The question is whether such a behaviour is possible.
This is where new information from cosmology comes in.
The information that we need is that the universe is flat and that
it has a positive cosmological constant.

Is there also direct information on cosmic topology?
In principle yes, since in a topologically nontrivial universe there should
be multiple images of objects in the sky. In practice this is difficult 
and one tries to use the cosmic microwave background (circles in the sky).
There is no convincing evidence that topology is present. So should we therefore ignore
this possibility? This is where the flatness of the universe comes in. In a curved
space there is a maximal scale where topology should appear, in a flat space there is none.
It could in principle be beyond the range that we can measure.
Flat and possibly anisotropic spaces are called Bianch-I type universes.
Given the fact that non-trivial topology is possible in a flat space, the next question is
whether the desirable dynamical behaviour is possible or likely. Namely the early universe
should be highly anisotropic, one dimension being much smaller than the others,
but at late times the universe should become isotropic as seen today.
This is where the positive cosmological constant comes in. Late time isotropisation
of the universe is actually a generic behaviour for  universes with a positive 
cosmological constant\cite{wald}.
The first example of such a universe is the Kasner solution\cite{kasner}.
At small times the solution looks like
$$ds^2 \cong -dt^2 + dx^2 + dy^2 +t^2 dz^2$$
Therefore the third dimension  gets compactified to zero at early times.
If this were the full solution at all times, it would be Minkowski space.
One can embed this solution in a full cosmology, where then at late times
one finds isotropy.

Therefore we suggest that the topology of the universe is $M_3 \times S_1$.
The radius of $S_1$ may be too large to see the topology at the present time.
However a remnant of the topology is the existence of a preferred direction
in the universe, that might be visible. Indeed there appears to be an allignment
of low multipoles along a preferred axis in the CMB data\cite{wmap}. Apparently this can be
explained in an inflationary Bianchi-I universe\cite{CMBan1,CMBan2}.

\section{The argument}
With these arguments I hope I have convinced you that three-dimensional dynamics
may be important for the four-dimensional world. The question is what does 
three dimensional physics look like. As an example one can take Yang-Mills theory.
Beyond the ordinary Yang-Mills term in the Lagrangian, with a gauge coupling $g$
 there is a second parity violating Chern Simons term with a mass 
$m$\cite{siegel,schonfeld,jackiw1,jackiw2,jackiw3}. 
The theory 
describes massive vectorbosons with only one polarization. The theory is invariant
under small gauge transformations, but the action gets shifted by a constant for
large (topologically non-trivial) transformations. Since the Lagrangian action appears
with an imaginary part in the path-integral, one will only have a full gauge invariance
when there is a quantization condition. One finds, that
the Yang-Mills charge $q_{YM} = \frac {4\pi m}{g^2}$
must be an integer.
What makes the quantization condition particularly interesting is that the Yang-Mills charge
gets renormalized\cite{rao}. There are corrections to the Yang-Mills charge through fermion
and vectorboson loops.
As a consequence one can derive the condition, that there must be an even number of fermions
in the fundamental representation\cite{witten,alvarez,redlich1,redlich2}.
For the connection with four-dimensional physics see\cite{frans1,frans2}.

For the case of Yang-Mills  and Maxwell fields the theory is well known in the literature.
Somewhat less well known is that also for gravitational fields a similar argument exists.
Only a few papers\cite{goni,vuorio,ojima,pisarski,yang,lerda}
 dealing with induced Chern-Simons terms exist in the literature.
In three-dimensional gravity there exists the Einstein Lagrangian and a Chern-Simons term.
Also here there is a quantization condition for the gravitational Chern-Simons charge.

The gravitational action in three dimensions contains two terms.
One is the ordinary Einstein Lagrangian:
\beq
L = -(1/\kappa^2)\sqrt{g} R
\eeq
where as usual, R is the curvature scalar, $g_{\mu\nu}$ is the metric tensor,
$g$ the determinant of the metric and $\kappa^2$ is Newton's constant.
To this action a Chern-Simons term can be added:
\beq
L_{CS} = -\frac{i}{4\kappa^2\mu}\epsilon^{\mu\nu\lambda}
(R_{\mu\nu ab} \omega_{\lambda}^{ab} + \frac{2}{3}\omega_{\mu a}^b
\omega_{\nu b}^c \omega_{\lambda c}^a).
\eeq
where
\beq
R_{\mu\nu ab}  = \partial_{\mu} \omega_{\nu ab}
+\omega_{\mu a}^c \omega_{\nu cb} -(\mu \leftrightarrow \nu)
\eeq
is the curvature tensor and $\omega_{\mu ab}$ is the spin connection.
The gravitational Chern-Simons charge
\beq
q_{gr} = \frac {6 \pi}{\mu \kappa^2}
\eeq
is quantized and has to be an integer.
The presence of matter fields however, fermions and vector bosons
with a Chern-Simons term, gives rise to an extra effective contribution
to the Chern-Simons charge $q_{gr}$.
\beq
q^{ren}_{gr} = q_{gr}
+\frac{1}{8} N_{g}~sign(m_{g}) -\frac{1}{16} N_f~sign(m_f)
\eeq
where $N_{g}$ is the number of vector bosons with 
topological mass $m_{g}$ and $N_f$ is the number of 
fermions of mass $m_f$. 

  It is important that the corrections
are only dependent on the sign of the mass and not its absolute value.
This means that also at zero mass an effect is present. Within the purely
three dimensional case one speaks therefore of a parity anomaly, since
the basic tree level Lagrangian does not violate parity. Embedding the theory in
four dimensions with a preferred direction it is easy to understand that the sign is
important, since the sign of the mass in the Chern-Simons like term is fixed when
one chooses an orientation for the coordinate basis vectors.
We now assume that the fundamental  gravitational
laws have no preferred direction, implying $q_{gr}=0$.
The meaning of this condition is that the gravitational field equations are given
by Einsteins equations, so that any early anisotropy comes from
the initial conditions and not from the equations. The assumption is technically
natural, since imposing it enhances the symmetry of the Lagrangian.
The complete effective Chern-Simons term is then induced by the matter fields.
In this case the quantization condition gives rise to the following
identity
\begin{equation}
N_f \mp 2 N_g = 0~~{\rm mod} (16)
\end{equation}
whereby the minus sign is to be taken when the fermions and the bosons have the
same sign of the mass.
It is assumed that the fermions separately and the bosons separately have the same sign
for the mass, which is a reasonable assumption when they are part of
the same multiplets in a unified theory, since otherwise one would break the
gauge symmetry.
We see that the condition (6) is fulfilled for the vector bosons by themselves
 if the gauge group is $SU(5)$, giving $N_g=24$ and also for  the fermions
by themselves, when they
are in the $16$-dimensional spinor representation of $SO(10)$.
Moreover
it is desirable that the effective renormalized gravitational Chern-Simons
charge $q^{ren}_{gr}=0$, since otherwise it is difficult to understand
that the late universe is even approximately isotropic, because the gravitational
field equations themselves would have a preferred direction.
This condition is fulfilled if there are three generations of fermions
$3\times 16 - 2\times 24 = 0$.\\
Though  at first sight these conditions look rather insignificant, they imply
 strong constraints, when one makes the assumptions that the
fermions should be in an automatically triangle anomaly free group and that they should
be in a fundamental representation. The solution appears to be essentially unique.

\section{Discussion}
It is to be remarked that the argument  contains little speculation.
On the cosmology part the Ansatz is more conservative than the standard
Robertson-Walker Ansatz that implies isotropy always. The unification with the
given
$SU(5)$ and $SO(10)$  multiplets is the absolute minimum that is possible within
grand unified theories.
The gravitational anomaly argument is  established mathematical physics.

On the speculative side one can wonder about symmetry breaking, other compactifications
and extra conditions. Taking the anomaly conditions into account a breaking
pattern $SU(5) \rightarrow SU(3)\times SU(2)\times U(1)$ appears to be preferred.
For this school probably the most relevant is the question
into the relation with quantum geometry.
From the argument it appears that a unique structure of gauge particles is needed
in order to have gravity in the theory. The argument is basically topological
and rather robust regarding the precise nature of quantum gravity. The argument 
works quite natural within the context of a quantization following the Einstein
equations. There appears to be no need for string theory. Therefore maybe the canonical way
towards quantum gravity is the right way. Naturally the results cry out for an underlying structure.
I have not been able to come to a conclusion here. I made some attempts using
octonions because of the factors of eight in the formulas, but found no convincing connection 
sofar.
Finally we are in a position to give tentative answers to the following
questions.\\

\noindent Rabi's question: Who ordered that?\\
Answer: the early universe.\\

\noindent Einstein's question: did God have a choice?\\
Answer: No, because He has to use perfect symmetry.\\

However the devil may have had something to do with the Higgs sector.
Maybe  one should  therefore,
 when one mentions the Higgs sector, not talk about
the God particle (L. Lederman), but about the devil's field.
More probably one should do neither.

\section{Acknowledgements}
I wish to thank to thank the organizers of the school and in particular
George Zoupanos for their hospitality.
This work was supported by the EU through the network HEPTOOLS.


\end{document}